# TM-vector: A Novel Forecasting Approach for Market stock movement with a Rich Representation of Twitter and Market data

A PREPRINT

**Faraz Sasani**
Economics and management Science (MEMS)
Humboldt University of Berlin
sasanifa@hu-berlin.de

**Ramin Mousa***
Department of Computer Science
Computer, Zanjan University
Raminmousa@znu.ac.ir

**Ali Karkehabadi**
Department of computer Science
Technology, UC Davis
Alikarkeh.abadi@gmail.com

Samin Dehbashi
Department of computer Science
Technology, UC Davis
samindehbashi@gmail.com

**Ali Mohammadi**
Department of Industrial Engineering
Sharif University of Technology, Tehran, Iran
alimohammadi@alum.sharif.edu

## ABSTRACT

Stock market forecasting has been a challenging part for many analysts and researchers. Trend analysis, statistical techniques, and movement indicators have traditionally been used to predict stock price movements, but text extraction has emerged as a promising method in recent years. The use of neural networks, especially recurrent neural networks, is abundant in the literature. In most studies, the impact of different users was considered equal or ignored, whereas users can have other effects. In the current study, we will introduce TM-vector[1] and then use this vector to train an IndRNN and ultimately model the market users' behavior. In the proposed model, TM-vector is simultaneously trained with both the extracted Twitter features and market information. Various factors have been used for the effectiveness of the proposed forecasting approach, including the characteristics of each individual user, their impact on each other, and their impact on the market, to predict market direction more accurately. Dow Jones 30 index has been used in current work. The accuracy obtained for predicting daily stock changes of Apple is based on various models, closed to over 95% and for the other stocks is significant. Our results indicate the effectiveness of TM-vector in predicting stock market direction.

## 1 Introduction

Today with the global rise of the stock market, the need for decision-making using stock market forecasting models has increased (1). Current stock market forecasting models still have low accuracy (2). Forecasting of the stock market using Recurrent neural networks (RNNs), as deep learning methods, has been investigated in the majority of articles. They reported that deploying these networks can lead to higher accuracy than other methods while using historical market statistics. (3) (4) (5) (6) (7) (8) (9) (10) (11) (12). In addition, many articles have examined the impact of tweets sentiments on predicting stock market changes. They ascertained that market changes have been affected by user behavior based on published tweets (13) (14) (15) (16) (17). Therefore, various analyzes of users' social behavior have been presented in this area, each of these analyzes can be effective in stock market prediction. The first is to analyze the sentiment of stock exchange and users. Researchers have proposed several approaches in recent years, all of which seek to improve the accuracy of emotion analysis that can adequately reflect user's behavior and stock news (18).
Another analysis is related to the social status of users in the network. This issue has received very little attention and has only been limited to select the user effectively. In other words, first, a small group of users is selected, and then, only their opinions are used in the prediction. Since all user reviews don't have the same impact and vary with each other, people's opinions can be crucial in predicting the social importance of individual changes over time. An additional problem with the accuracy of deep learning methods pertains to the volume of data input to deep networks, which can significantly affect prediction accuracy. However, all the available defined methods are able to make daily forecasts on the stock exchange using an input sample of the number of days that examine the market, which this input is limited to the past few years. (19)(20).

Given the background of the available approaches and the new needs and capabilities that deep learning has brought to this field, we have proposed a new approach that can be applied to all similar areas. The main contribution of this research is to consider each user in the form of a vector, which can incorporate all of the effective predictive features and then be delivered to the deep network as an input sample. In fact, this approach enables us to reflect the behavior of users in the market, while, the relationships between users extracted using the Independently Recurrent Neural Network (IndRNN).

The results showed that the proposed TM-vector approach has been able to improve the predictions at low cost and effectively alleviate the problem of lack of training data in the deep network. In this study, Twitter users were used to model the behavior of market users, using a data collection and preparation process in several steps. First, all the tweets related to the required stocks are collected by special hashtags, and then the social attributes that influence the value of each user and the sentiment of feedback are extracted. It is attempted to calculate new features to estimate the success rate of the user in forecasting, then using these features, the user vector is constructed and sent to the appropriate network, so as to be able to forecast daily stocks in the Dow Jones 30.

## 2 Related work

The sentiment analysis-based, and Technical-based approaches have been used in many types of research to solve the problem of financial time series analysis. This section provides a brief review of some of the works in the literature that have used these techniques

### 2.1 Sentiment analysis-based stock prediction

Sentiment Analysis (SA) is an NLP task that attempts to classify a document into a set of polar classes (negative, positive, or neutral) (21). Generally, SA is intended to determine the attitude of a speaker or a writer in relation to a subject, which is generally expressed as a non-structural text (22). SA has become an important part of the stock market, as the idea of SA based on different data sources can provide insights into how stock markets respond to different types of news in the short and medium time. (23) proposed a method that measures sentiment from sources or the sentiment be- hind the news to determine its impact on stock markets also, Lee et al. (24) presented an approach to determine the importance of SA in stock market forecasting. The authors developed a system to predict whether stock prices will stay high or low by SA in the Form 8-K reports of their respective stocks. Derakhshan et al. (25) implemented "LDA-POS" method which addressed the issue by incorporating part-of- speech tags into methods of topic modeling. The accuracy of the model for both Persian and English datasets report 55.33% and 56.24% respectively.
In the domain of stock market prediction, the main ap- plication of SA is prediction of the emotion of people that has effect on stock price. Batra et al. (26) implemented a Support Vector Machine (SVM) model for forecasting and analyzing stock tweets which are related to Apple products. They found a positive relation between people's viewpoint and stock price market data with accuracy of 76.65%. More- over, SA helps investors to make acute decisions.

Ren et al. in 2018 (27)implemented a model based on combination of investors' sentiment and machine learning methods with accuracy of 89.93%. SA approach also adds the value to stock market through prediction in sales of merchandises. Pai et al. (28) presented a useful framework with combination of hybrid multivariate regression with time series pre- diction models. The result of this method on vehicle sales data illustrates that an accurate forecasting in comparison to other prediction models. One of the main applications of SA in stock market is financial decision. Das et al. (29) applied Lambda Architecture for data processing. They acquired ac- curate results were deployed users' behavior by sentiment analysis through LSTM algorithm. Their proposed model consisted of two basic phases. In the first phase, they attempted to classify the tweets as positive and negative using NLP. In the second phase, machine learning algorithms such as Decision Tree (DT), SVM, and Naive Bayes (NB) were used to classify polar tweets.

### 2.2 Technical-based stock prediction

Technical analysis is the study of financial market movements in financial areas. Technicians are usually the ones who perform these techniques. They use past price and volume data, which is graphically displayed, to decide where future financial markets will move (30). Technicians believe that all stock information is reflected in the stock price. Over the years, technicians have used a variety of methods to dis- play price and volume data to help simplify decision tips. These tips are referred to as technical indicators. Various approaches have been proposed in the literature for this purpose. Most of these approaches are based on deep learning systems. Deep learning has achieved acceptable results due to the automatic selection of features and the lack of need for manual selection.

Various technical indicators such as RSI (Relative Strength Index), CCI (Commodity Channel Index), Moving Averages, and etc. have been used for this purpose, which are detailed in Section 3.1.5. Bernal et al (31) proposed a recursive network called Echo State Networks (ESN) to predict S&P 500 stock using features such as price, moving average, and volume. This network was tested on 50 stocks and was able to get a test error of 0.0027. Balling et al (32) tested ensemble approaches, including Random Forest, AdaBoost, and Kernel Factory against single classifier models such as Neural Networks, Logistic Regression, SVM, and K-Nearest Neighbor on data from 5767 European companies, which gave Random Forest a better result than other approaches. Other supervised approaches such as Neural Network (NN)[45], Multiple Linear Regression (LMLR)(33), Supper Vector Regression (SVR)(34), j48 algorithm (35), LSTM and DNN (36), and RNN(LSTM) (37) have also been used in the literature to extract and introduce technical indicators.

Diverse combined indicators and requirement analysis are also provided in the literatures. They have not considered the impact of users in these predictions or have used it as a constant factor for all users. The Table 1 shows the major combinations of different features in the available re- searches that show the difference between our work and that of others. In (38), KSE-100 stocks were analyzed. In this approach, the issue was addressed as a classification problem. The use of simple machine learning approaches is one of the strengths of this research, which despite the simplicity of the models, has reached acceptable results, but the selection of features manually is one of the weaknesses of this research. Research (39) has been a little more detailed in classification and has used three strategies (1) (sell), 0 (hold), and - 1 (buy) for classification. This study considered the use of deep learning models to extract features automatically. Not mentioning the number of samples in both studies (38) and (39) makes it somewhat more difficult to review the results, and only the use of the Accuracy criterion increases the possibility of overfitting the model. In (40), feature-based solutions, fuzzy clustering, and fuzzy NN were used for price prediction. In this study, the problem was treated as a regression that aimed to predict the US SSP 500 index price for 361 days. Moreover, the RMSE criterion was used.

The strength of this research was the use of fuzzy function with NN, which despite the small number of samples, was able to achieve a lower RMSE than other approaches such as ARIMA, Linear Regression and SVR. In (41), the authors examined the market direction (up or down) on Korea KOSPI200; in this study, the classification problem was considered, and the ANN approach, along with feature selection as their proposed solution, was able to reach an acceptable value. One of the strengths of this research is to evaluate the models with different statistical tests. Research (42)(44)(45)(46)(49) are slightly different from other research because, in these re- searches, the market trend was considered as a goal and template matching patterns were used for the market trend. In this matter, classification or regression is not considered, and only pattern extraction is considered. Other categories of discounts (43)(47)(48)(50) have a regression view of the problems and consider the price or return. In these approaches, regression evaluation criteria such as NMSE, RMSE, MAE, and MI to evaluate the models used (49).

Table 1: A summery of recent studies on stock market prediction.

| *Reference* | Data Type | Target Output | Samples | Method | Performance |
|---|---|---|---|---|---|
| (38) | KSE-100 index | market trend (bull/bear-flag) | - | feature selection+ Machine learning algorithms | Accuracy |
| | Dow Jones | Trading strategy (1 (sell),0 (hold),-1(buy)) | - | feature selection+ deep learning | Accuracy |
| (39) | | | | | |
| (40) | US S&SP 500 index | stock price | 361 | feature selection+ fuzzy | RMSE |
| (41) | Korea KOSPI200 | market direction (up or down) | *3,650* | feature selection+ ANN | statistical tests |
| (42) | US Dow Jones | market trend (bull/bear-flag) | 91,307 | template matching | trading simulation |
| (43) | India CNX and BSE indices | stock price | 2,393 | SVR+ ANN, RF, SVR | MAPE, MAE, RMSE, MSE |
| (44) | Taiwan TAIEXa and US NASDAQ | market trend | 3,818 | dimension reduction +template matching | trading simulation |
| (45) | World 22 stock market indices | trading signal | 756 | particle swarm optimization+ ANN | trading simulation |
| (46) | Greece ASE general index | portfolio composition | 3,907 | fuzzy system | trading simulation |
| (47) | Japan Nikkei 225 index | stock return | 237 | ANN +genetic algorithm, simulated annealing | MSE |
| (48) | US Apple stock | stock price | 19,109 | deep NN | MSE, directional accuracy |
| (49) | US SPDR S&P | market direction (up or down) | 2,518 | dimension reduction +ANN | trading simulation, statistical tests |
| (50) | Korea KOSPI 38 stock returns | stock return | 73,041 | data representation+ deep NN | NMSE, RMSE, MAE, MI |
| (51) | EUR/USD | sentiment analysis | 25,793 | text and image representation + CapsuleNet | Accuracy, Precision, Recall, F1 |
| Our study | Dow 30 | market direction (up or down) | 1,589,185 | feature representation+ IndRNN | Accuracy, Precision, Recall, F1 |

The study (51) considered the problem as emotion analysis and used the new capsule deep learning approach. The strength of this approach is the use of visual and textual information in EUR/USD in a single model called TI-Capsule was. The weaknesses of the studied approaches can be summarized in the following four cases:

- Not considering social network information and only using market information increases the possibility of underfitting and overfitting the model.((38) (39) (40) (41) (43) (44) (45) (46) (47) (49) (50)).
- The use of classification allows entering into transactions that do not bring significant profits. For example, if the stock price the next day is $ 100 and today's price is $ 99.80, it will be assigned the up, bullish, bull, or buy label.((38)(39)(49),and TM-vector).
- Manual feature Selection is time-consuming and challenging because the model becomes highly dependent on the feature ((38)(41)).
- Combines text, market, and image information with time overhead and memory consumption. (51) and TM-vector)

## 3 Methodology

The main contribution of this work is introducing TM- Vector, which can be helpful in any area where user behavior modeling is effective in predicting the trend of that area like the stock exchange. The TM-Vector, provides a full representation of user's behavior and market behavior in vector space. In any areas that the user behavior, lack of adequate supplies, and demands dominate the current trend, this idea can be used to predict the future trend of that area. TM-vector in stock market prediction in which transformations are affected by user behavior, can be used by Twitter users to represent market investors and create a vector for each of them. The user vector can be composed of different parts, each of which adds new features to the prediction model.

In this work, we proposed the five types of features to create a user vector, and then we evaluated the prediction matrix by accuracy. Then the best category of features, the best structure of the deep neural network and the most accurate prediction label based on the results obtained, are selected as the best model.

Another practical factor which is considered in each user's vector, is the use of market data as a part of the user's knowledge in investment decision making. Moreover, this feature has been able to make a significant improvement in the TM- vector compared to previous models where each input rep- resented one day instead of one user.

## 3.1 TM-vector components

The market data is Dow30 in the form of 'Open', 'High', 'Low', 'Close', 'Adj Close' from 22-09-2021 to 02-02-2022. This data has been collected and completed in several stages for each stock. At each stage more information and feature are added to the data. The data is stored in separate rows where each row can be considered an input sample for model training or testing. The main purpose of introducing TM-Vector is to provide a comprehensive and effective representation of stock market data together with the user data (from twitter). It dis- plays various information, including:

1. Tweet text
2. Social network analysis
3. Sentiment of tweet text
4. User History
5. Stock market data

Figure 1 shows an example of a TM-Vector resulting from the collection and extraction of 5 categories of influential user characteristics and market data indicators in market direction prediction. Each of these five categories will be discussed in more details in the following subsections.

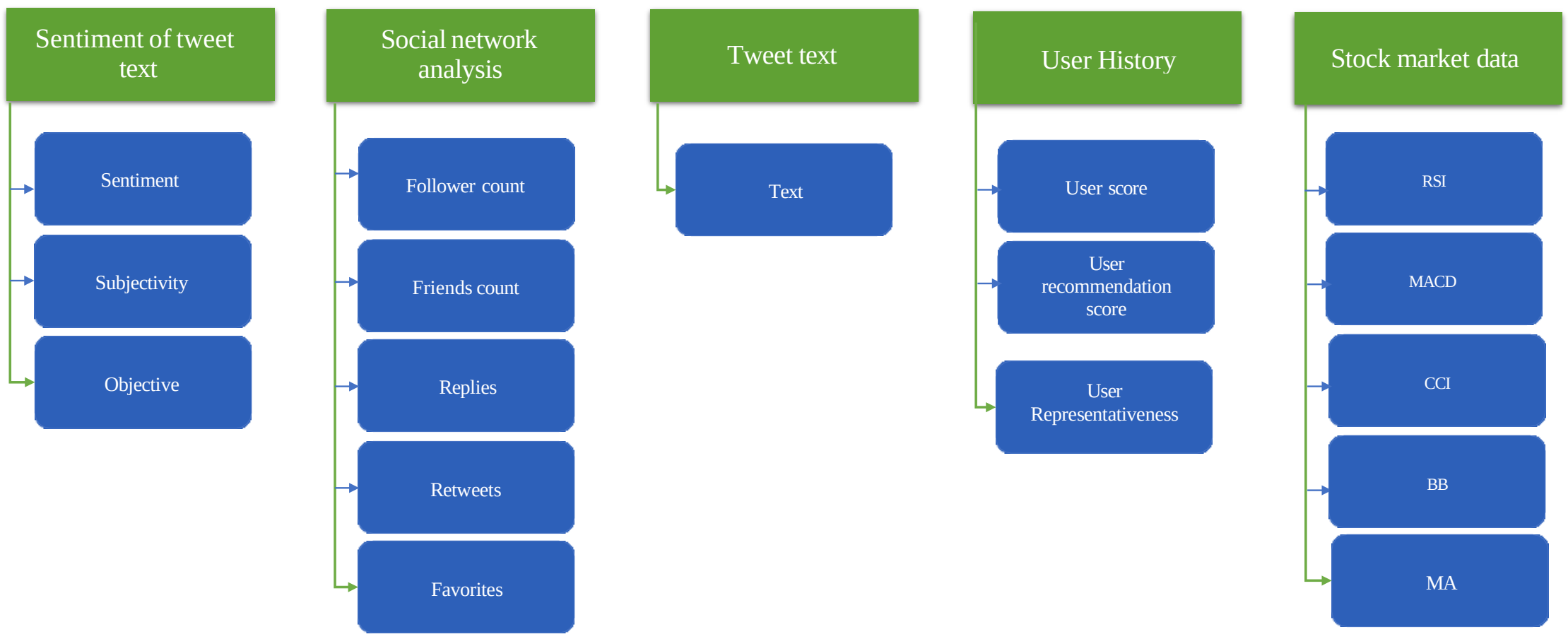

Figure 1: Categories of influential user characteristics

### 3.1.1 Tweet Text

One of the most important information that can be a good representative of user influences, is the text or content of the tweets which users publish in the network. To provide a vector representation of tweet text, firstly, each tweet is tokenized into sequence of its constituent words. Then all of the stop-words and punctuation are removed from this sequence. This removing is important as it will lead to some improvements in overall performance of the training algorithm and makes the classification model learn better features. After removing these items, the input sequence is map to a sequence of low-dimensional distributed representations for each word, using pre-trained word vectors. After mapping, the sequence of these words can be displayed as that $s = [X^1, X^2, ..., X^n]$ that $X^i \epsilon R^k$ is the k-dimensional word vector corresponding to the $i - th$ word in the sequence. One of the limitations of deep learning models is that the input vectors' dimension must be equal, while they are different, due to variable length of different tweet text. To solve this problem, generally, a threshold is used. All the inputs padded with a value (if less than the threshold) or cropped (if higher than the threshold). The maximum sequence length in the dataset is considered as the threshold for addressing this problem and all sequences that are less than the thresh- old padded with zero. The maximum length is used so as to prevent cropping, which generally may delete some useful data. The sequence of length n (padded where necessary) is represented as:

$$X^{1:n} = X^1 \oplus X^2 \oplus X^3 \oplus \ldots \oplus X^n$$

where $\oplus$ is the concatenation operator and $X^{1:n}$ is called **tweet vector**.

### 3.1.2 Social Network Analysis

In this section we extract several social influential features of the user that can result in successful prediction. More active users are more likely to be effective because they are more likely to influence other users, and therefore have to be considered more. Several features can determine the amount of user activity. Mainly, the number his/her followers, the number of users he/she follows, the number of times the user tweets are re-tweeted, reply, and favorites. These features are essential features that can be used to reflect user's activeness in a network and is represented by a vector which is called a **User vector**.

### 3.1.3 Sentiment of Tweet text

The sentiment of a tweet can be one of the essential features related to a tweet text which is published by a user and reflects the general attitude of a user about the posted text in a social network (e.g., twitter). This feature can be expressed in terms of three different properties/tags of Objective, Subjective and Sentiment. Using these three tags a numerical sentiment vector is generated, in which the Subjective and Objective properties are set as numbers between 0 and 1, while the Sentiment property rated as -1 (negative), 0 (neutral) and 1 (positive). The TextBlob[1] tool is used to generate the values of the three properties that then lead to a sentiment vector.

### 3.1.4 User History

There are a number of features specific to each user that can have a direct impact on forecasts. These features mainly related to the history of a user activity and fame of him/her in a social network. The followings are the features can be extracted per user from a social network.

1. User score: The score aims to reflect the percentage of successful prediction of each user with respect to their tweets. To compute this score, firstly the sentiment of user's texts (user's opinion about each stock) is extracted as mentioned in the section 4.1.3. Then using the actual information about the stock predictions in the gold dataset, we can determine how successful the user was in predicting each stock. Let $\bar{y}^t$ be the prediction of user *a* at time *t* and consider neutral comments 1 and also $\bar{y}_a^t$ be the actual values of the stock tag of user *a* at time *t*, then we can rate each tweet of user *a* at time *t* by the following formula:

$$TweetScore_a^t = \begin{cases} 1, if\ y_a^t == \bar{y}_a^t \\ -1, if\ y_a^t! = \bar{y}_a^t \end{cases} \quad (1)$$

   Given the Tweet score values up at time t for each user tweet, the correct and incorrect prediction score of each user can be obtained at time t through the following relationships:

$$PUS_a^t = |\sum_{t=start}^{T-1} TS_a^t|(TU = a, r_{at} = 1) \quad (2)$$

$$NUS_a^t = |\sum_{t=start}^{T-1} TS_a^t|(TU = a, r_{at} = -1) \quad (3)$$

   In this respect, $t$ is the time interval from the start of data collection to the current record, $PUS_a^t$ is the Positive User Score, $NUS_a^t$ is the Negative User Score, $TS$ is the Tweet Score, and $TU$ is the Tweet User, rat is the rating of tweet score which is 1/-1 for positive/negative user predictions per a tweet based on formula 2.

2. User recommendation score: Let $AR$ be the Average Rating of the Author until time $t$, which is obtained using the following equation:

$$AR_a^t = \frac{PUS_a^t}{H_a^t} \quad (4)$$

[1] https://textblob.readthedocs.io/en/dev/

Where $PUS_a^t$ the number of successful user predictions which is obtained from equation 2, and $H_a^t$ is the sum of the number of comments an author has published until time t. Given these, it can be said that the User Recommendation Score $URS_a^t$ can be computed as the following relationship:

$$URS_a^t = \begin{cases} 0, if\ AR_a^t = 0 \\ otherwise\ 1 + \log(AC_m^i) \end{cases} \quad (5)$$

3. User Representativeness: Using the $AR_a^t$ and $URS_a^t$ for each user $a$ at time *t*, the representativeness of User ($RA_a^t$) value is obtained by the following equation:

$$RA_a^t = \frac{AR_a^t + PUS_a^t}{2} \qquad (6)$$

### 3.1.5 market data

Market data includes the values related to historical stock data including: open, high, low, close, and Adj close. Some important indicators are computed using these concepts which will be explained in details below.

#### 3.1.5.1 Moving Averages (MA)

Moving average is one of the main tools used to analyze financial time series. In short, the moving average is the average of the simple weighted sum (average) calculated within the selected historical price range. Financial data is usually noisy, if we want to show today's price as the average of today and 2 days ago, all the ups and downs are done on average. By using more historical prices (increase period), we can reach a smoother price that shows the trend, despite price fluctuations (56).

If $x_i$is defined as a value at time *i*, the moving average at time *t* can be calculated as follows [5]:

$$MA_t^n = \frac{1}{n}\sum_{i=1}^{n} x_{i-1} \qquad (7)$$

Moving averages are used to show the trend of noisy data, and also produce smoother values. Smoothing often improves the consistency of conclusions and predictions as well.

#### 3.1.5.2 Relative Strength Index (RSI)

RSI is one of the most well known and widely used technical analysis indices (52). RSI is a moving index that measures the amount of recent price changes to check the "buy" or "sell" saturation conditions at the price of a stock or other financial instrument. The RSI is represented as an oscillator (a line graph that moves between two limits) and can fluctuate between 0 and 100 (53). The traditional interpretation of this oscillator is that values above 70 indicate the buying situation zone, so that the asset is priced higher than the intrinsic value of the stock and may be prepared to correct the trend, while RSI below 30 indicates that the stock market value is lower than its intrinsic value. To determine the RSI, according to formulas 8 and 9, the closing price increase (upward change) (U) or the closing price decrease (downward change) (D) is calculated for each day (53).

$$U_{close} = close_{today} - close_{yesterday} \quad (8)$$

$$D_{close} = close_{yesterday} - close_{today} \quad (9)$$

If *U* is positive for a particular day, *D* is replaced by 0 for that day, and vice versa, if *D* is positive for a particular day, *U* is replaced by 0 for the corresponding day. To calculate the RSI, the exponential moving average (EMA) for *U* and for *D* is determined using the "coefficient" ($a$) based on the number of specified days (N). The moving average is used to limit the impact of random factors that occur on the unusual mean.

$$a = \frac{2}{N+1} \quad (10)$$

The EMA uses the "$a$" coefficient to give different weight of importance to the data considered in the calculation, depending on their time. Therefore, older data weigh less in the EMA and newer data weigh more. According to formula 10, the "multiplication" of is determined in relation to the number of days *N* (*N* = 27).

$$SMA_n = \frac{X_1+X_2+\cdots+X_N}{N} \quad (11)$$

According to formula 11, determining the EMA requires calculating the Simple Moving Average (SMA) of the data related to the first N Day in the sequence. The exponential moving average (EMA) N + 1 of the day is determined as follows:

$$EMA_{N+1} = a * X_{N+1} + (1-a) * SMA_n \quad (12)$$

After determining the exponential moving average, the relative resistance (RS) is obtained through the following equation (53):

$$RS = \frac{EMA\ of\ U}{EMA\ of\ D} \quad (13)$$

This value is converted into an index that can be between 0 and 100 units, according to formula (7) this index is called the relative resistance index (RSI).

$$RSI = 100 - 100 * \frac{1}{1+RS} \quad (14)$$

#### 3.1.5.3 MovingAverage Convergence Divergence (MACD)

The MACD is commonly used to measure the strength of price movements. This is done by measuring the difference between the exponential moving averages 12 and 26. MACD calculations are based on the moving average (EMA) of final prices. EMA was defined as Equation 12. Below are the MACD calculations (54):

$$MACD = \sum_{i=1}^{n} EMA_k(L) - \sum_{i=1}^{n} EMAd(I) \quad (15)$$

#### 3.1.5.4 Commodity Channel Index (CCI)

CCI is the most common technical tool for investors. The purpose of the CCI is to identify periodic trends and trend changes. CCI is an oscillator that oscillates without limits. Oscillators are built to check for oversold areas. The CCI calculation is presented below. CCI is an oscillation that measures the current price level relative to the average price over a period of time (55).Typical price is $\sum_{i=1}^{p} \frac{High+Low+Close}{3}$, P is the *number of periods,* MA is $\sum_{i=1}^{p} \frac{Typical\ Price}{P}$ and $mean\ devation = \sum_{i=1}^{P} \frac{Typical\ Price-MA}{P}$.

$$CCI = \frac{Typical\ Price-MA}{0.015*MeanDeviation} \quad (16)$$

The CCI trading system generates buy signals when the CCI line crosses -100 and generates sell signals when the CCI line crosses the 100.

#### 3.1.5.5 Bollinger Bands (BB)

The Bollinger Bands indicator is a graphical index of fluctuations. This index consists of upper and lower bands or bands that react to oscillating changes. When the volatility of a hypothetical currency pair is high, the distance between the two bands increases. When the fluctuations of a currency pair are low, the distance between the two bands also decreases and both begin to tighten. The BB are used for long term as well as short term analysis. The BB indicator is made up of three main components: the Upper Band (UB), the Middle Band (MB), and the Lower Band (LB). The formulas for BB are given below (57) where $n$ is smoothing period, $m$ is number of standard deviations, and $\sigma(Closingprice, n)$ standard deviation over last $n$ periods of closing price:

$$UB = [SMA(Closingprice, n)] + [m * \sigma(Closingprice, n)] \quad (17)$$

$$MB = nperiodAimpleMovingAverage(SMA) \quad (18)$$

$$LB = [SMA(Closingprice, n)] - [m * \sigma(Closingprice, n)] \quad (19)$$

# 4 Recurrent Neural Networks

Recurrent Neural Networks (RNNs) are a type of neural networks that have direct cycles between their units. These cycles allow the neural network to make the internal state of the network and add the concept of time to the model. Which allows the network to capture dynamic behavior. At time $t$ each network node receives two inputs from the current inputs $x^t$ and the hidden node values $h^{t-1}$. Based on these inputs, the output of the hidden layer $h^{(t)}$ is obtained from the following equation (58):

$$h^{(t)} = \sigma\left(W^{hx}x^{(t)} + W^{hh}h^{(t-1)} + b_h\right) \quad (20)$$

Here $W^{hx}$ is the matrix of weight between the input and the hidden layer, $W^{hh}$ is the matrix of recurrent weights between the hidden layer and itself at adjacent time steps. $\sigma$ is activation function, $b_h$ is bias parameter, $x^{(t)}$ is current input data, and $h^{(t-1)}$ is previous hidden layer output. The network output is obtained from the following equation:

$$y^{(t)} = \sigma\left(W^{yh}h^{(t)} + b_y\right) \quad (21)$$

Here $W^{yh}$is the matrix of weight between the input and the output layer, $\sigma$ is a non-linearity function that used for multiclass classification, $h^{(t)}$ is the output of the hidden layer which is obtained by the equation 20, and $b_y$ is bias parameter. The bias allows each node to learn an offset.

## 4.1 Independent Recurrent Neural Network

An independently recurrent neural network (IndRNN) can be defined as the following function (59):

$$h_t = \sigma(Wx_i + u \odot h_{t-1}) + b \quad (22)$$

Where $h_t \in R^N$ is the hidden state at time step t, $\sigma$ is sigmoid activation function,$W^{(N*M)}$ learnable matrix, $x_t \in R^M$ is input at time $t$, $u \in R^n$ is current imput, $\odot$ is the Hadamard product, and $b \in R^n$ the bias of the neurons. and M also refer to the number of neurons in the recurrent layer and the input dimensions respective. One of the most important differences between IndRNN and traditional RNNs is the return weight u, which is traditionally a matrix that transforms input by matrix multiplication, but in IndRNN this value is a vector and input through processes multiplication based on elements. In IndRNN, each neuron in each layer is independent of the other neurons, hence it is also called "independently recurrent"(59). In IndRNN for the $n$ - $th$ neuron, the hidden state $h_{(n,t)}$ can be obtained as:

$$h_{(n,t)} = \sigma\left(w_n X_t + u_n h_{(n,t-1)}\right) + b_n \quad (23)$$

## 4.2 Long-Short Term Memory (LSTM)

LSTM is a type of RNN architecture that addresses the vanishing gradient problem and allows learning of long-term dependencies. The initial idea of LSTM was proposed by German researchers, Hochreiter and Schmidhuber (60), to address the long-term dependency problem. These networks keep more information than the recurrent network in a cell. The information inside the cell can be read, written, and stored like computer memory. Each cell has four gates, namely the input gate i, the output gate o, the forget gate f, and the cell update gate g. At each time step, there are the following computations for each gate (61):

(24)

$$f^{<k>} = \sigma\left(W_f r^{<k>} + U_f h^{<k-1>} + b_f\right)$$
$$i^{<k>} = \sigma(W_i r^{<k>} + U_i h^{<k-1>} + b_i)$$
$$g^{<k>} = tanh\left(W_g r^{<k>} + U_g h^{<k-1>} + b_g\right)$$
$$o^{<k>} = \sigma(W_o r^{<k>} + U_o h^{<k-1>} + b_o)$$

Where $[W_i, W_f, W_g, W_o, b_i, b_f, b_g, b_o]$ is the set of parameters to be learned. $q^{<k>}$ is updated through the following relationship:

$$q^{<k>} = f^{<k>} \odot q^{<k-1>} + i^{<k>} \odot g^{<k>} \quad (25)$$

where the $o$ symbol represents the element-wise product be- tween two vectors. Finally, the activation of the cell is accomplished through the following relationship:

$$E = h^{<k>} = o^{<k>} \odot \tanh(q^{<k>}) \quad (26)$$

To determine the class in the classification task, it is sufficient to apply a sigmoid to the encoding phase output as follows:

$$z = sigmoid(W_z E + b_z) \quad (27)$$

### 4.3 Gated Recurrent Units (GRU)

A GRU is a variation of RNNs, mainly LSTM, without a separate memory cell. It was first designed by Kyunghyun Cho in his paper about Neural Machine Translation (62). Contrary to LSTM approach, this network has only two gates and has a relatively faster learning process. In GRU if the in- put matrix is represented by $Input = [X_1, X_2, \dots, X_n]$ then the $h_t = [h_1, h_2, \dots, h_t]$ that represent the hidden vectors sequence in GRU is calculated by the following equations:

(28)

$$z_t = \sigma\left(w_z x_t + U_z h_{(t-1)}\right) + b_z$$

$$r_t = \sigma\left(w_r x_r + U_r h_{(t-1)}\right) + b_r$$

$$h'_t = \sigma\left(w_h x_t + U_h(r_t \odot h_{(t-1)})\right) + b_h$$

$$h_t = (1 - z) h_{(t-1)} + z_t h'_t$$

Where $z_t$ is update gate, $r_t$is rest gate, $h'_t$ is candidate gate, and $h_t$ is output activation. $w_z, w_r, w_h, U_z, U_r, U_h$ are learnable matrixes, $b_z$, $b_r$, $b_h$ are learnable biases and $\sigma$ is sigmoid activation function, and $\odot$is an element-wise multiplication. GRUs in normal model will simulate input in one direction.

## 5 Learning of the TM-vector Model by IndRNN

To understand the impact of different features on the proposed Tm-vector, three different models are examined, including:

1. Text based IndRNN
2. Numerical based IndRNN
3. Text and numerical based IndRNN

The details of these three models are explained below:

- **Text based IndRNN**: In this model, the raw text (sequence of words) that have been transformed into a set of dense vectors is used as input. One of the most important issues that should be taken into account when working with the neural network on the text is that we can not directly feed the raw text to the neural network, as the neural network receives D-dimensional feature vectors. One-hot encoding is not suitable due to the length of the dictionary size. So, there is a need to embed feature into a D-dimensional space and represent it as a dense vector in the space. The solution is called word embedding, which creates a single dimensional dense vector for each word. The vectors are very flexible and help to avoid the curse of dimensionality. Generally, for creating dense vectors, the word embedding methods are trained on a large volume dataset. The pre-trained word2vec Google News vectors word embedding has been used in our approach. It trained on 100 billion words from Google News producing a vocabulary of 3 million words that are available at Github[2]. The IndRNN is used to learn word features. This model shows the effect of the tweet text, published by the users on the market prediction and is presented in Figure 2.

- **Numerical based IndRNN**: In this model, only the scaled numerical features are given to the IndRNN. In this model, each record of feature is displayed as a vector $x = [x^{<1>}, x^{<2>}, ..., x^{<n>}]$, where the $i - th$ index of this vector represents the $i - the$ property of the input component. The values are normalized due to the varying numerical scale of the input components. Linear Scale Transformation (Max-Min) [41] has been used in current work for this purpose. We calculate the normalized value for each index of the input component according to the following equation:

$$r_{ij} = \frac{(x_{ij} - x_j^{min})}{x_j^{max} - x_j^{min}} \quad (29)$$

The normalized value of the input component can be represented as $r = [r^{<1>}, r^{<2>}, ..., r^{<n>}]$ where each $r^{<i>}$ is the normalized value of $x^{<i>}$ of the input component. The purpose of is to investigate the effect of numerical features without considering textual features on market data. The data in this model are sequential which is feed to the model. In this model, the structure similar to model based on textual was considered. The model architecture is shown in Figure 3.

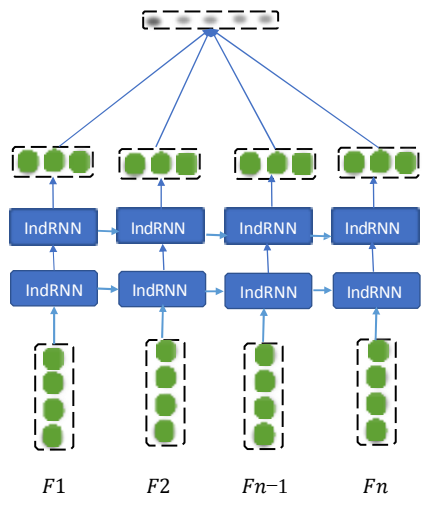

Figure 2: Text based IndRNN schema.

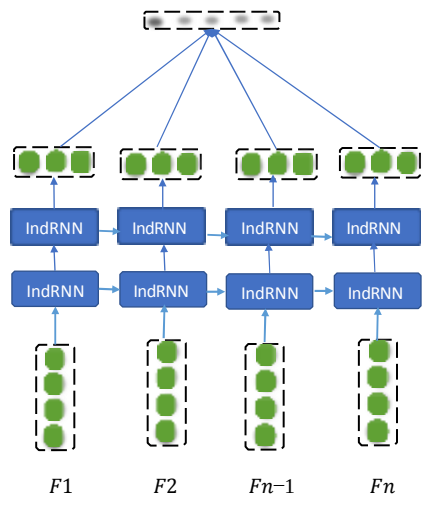

Figure 3: Numerical based IndRNN schema.

- **Text and numerical based IndRNN**: The purpose of text and numerical based IndRNN is to learn from combined numerical and text features. The model architecture is shown in Figure 4, which is the combination of two texts based and numerical based net- works, which connects the two network is parallel. The left-hand side network receives the textual data and extracts the text latent features from the text, and the right-hand side network receives numerical market data and extracts the latent features from them. Finally, the two latent properties are combined and as- signed to a fully connected layer for classification as shown in Figure 4.

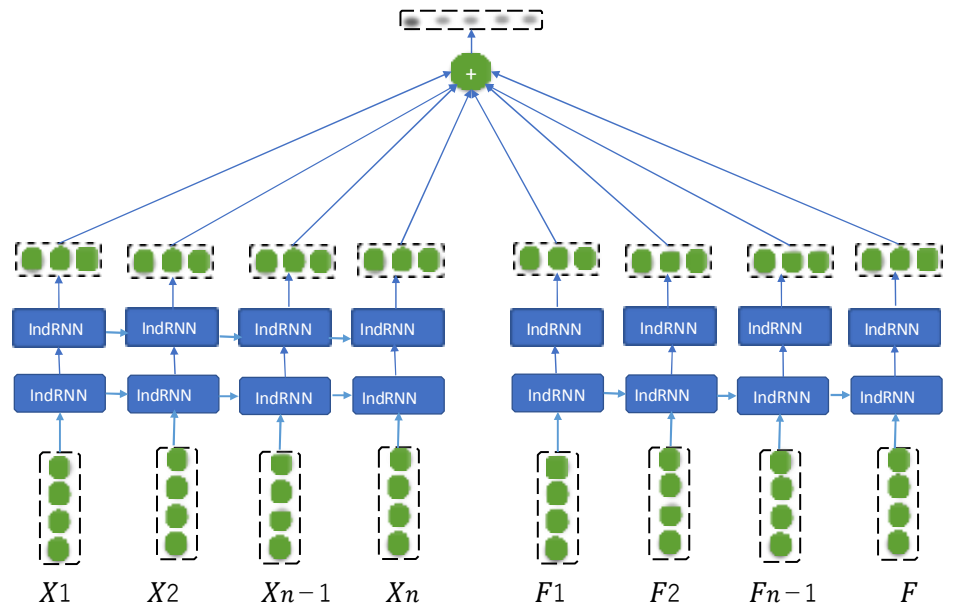

Figure 4: Text and numerical based IndRNN

## 6 Materials and evaluation protocols

This section provides a description of the data, extracted features and tools for collecting and preparing data, to be feed to the provided models. Each set of features is collected from different sources and incrementally added to previous data sample, and then their influential and essential features are extracted.

[2] https://github.com/mmihaltz/word2vec-GoogleNews-vectors

### 6.1 Yahoo finance app

Online trading and instant stock market statistics for New York and NASDAQ are provided by numerous free web sites and software tools, mainly Yahoo and Google. However, the statistics are publicly available with a 20-minute delay, but can be accessed by brokerage clients if data is needed without delay. For this purpose, using the data available on the Yahoo are site, the price information of each stock was extracted on specific days, from 22-09-2021 to 02-02-2022. This data is used in the labeling section for model training. It calculates the increase or decrease in the price of each share using real data and is used as the correct label for model training as well as the evaluation of the trained model. To calculate the increase or decrease of the price, the end or starting price of each day is compared with the corresponding price of the next day. If today's price is higher than tomorrow's price, label 0 is assigned; otherwise, label 1 is assigned. Data collection from the Yahoo site can be easily performed within a short time using the Yahoofinance[3] tool. Table 2 shows some sample records of the data extracted for Apple stock.

Table 2 :Sample of labeled data.

| Date | Open | High | Low | Close | Adj Close | Label |
|---|---|---|---|---|---|---|
| 02/05/2020 | 175.35 | 175.81 | 173.93 | 175.35 | 171.0376 | 0 |
| 03/05/2020 | 175.55 | 175.69 | 173.41 | 175.33 | 171.018 | 1 |
| 04/05/2020 | 174.16 | 175.3 | 173 | 174.29 | 171.9788 | 0 |
| 05/05/2020 | 175.59 | 177.49 | 174.49 | 175.28 | 170.95 | 1 |
| 06/05/2020 | 176.18 | 178.36 | 175.65 | 177.09 | 174.7256 | 1 |
| 07/05/2020 | 176.9 | 178.40 | 176.14 | 176.19 | 175.8427 | 0 |

### 6.2 Twitter API

Twitter is often known as a public platform, and various APIs have been introduced to collect its data [2]. In this work, we used GetOld Tweets[4], Tweepy[5], and Textblob[6] to collect data and create new features. Some tools provide access to older tweets, and others have some download restrictions. GetOldTweets3 is a completely free Twitter data gathering tool that also supports hybrid search and word search features, allowing to access older tweets. This API has very useful information like id (str), permalink (str), username (str), to (str), text (str), date (Date Time) in coordinated universal time (or UTC), retweets (int), favorites (int), mentions (str), hashtags (str), and geo (str). The features that GetOld Tweets extracts are useful but do not contain crucial social information such as the number of followers and followings, so we use Tweepy to extract some other useful social features for each user. This powerful tool is also used to collect tweeter data that uses the OAuth mechanism for authentication. The next tool is a Textblob that can extract the emotional tag of each tweet's text. It also uses the OAuth authentication mechanism, just like the previous one.

### 6.3 Statistical analysis of collected tweets

Table 3 shows the number of samples collected per share, hashtags used, positive rate, negative rate, and unique user of the data. Due to the difference in the number of tweets per day for each stock, the various number of tweets collected per for each stock.

### 6.4 EVALUATION PROTOCOLS

The four metrics of accuracy (ACC), Precision, Recall, and F1 are used as evaluation criteria for experimental results, which are defined as follows:

$$Precision = \frac{TP}{TP+FP} \quad (30)$$

[3] https://pypi.org/project/yahoo-finance/
[4] https://pypi.org/project/GetOldTweets3/
[5] http://docs.tweepy.org/en/latest/
[6] https://textblob.readthedocs.io/en/dev/

$$Recall = \frac{TP}{TP + FN} \quad (31)$$

$$F1 = 2 * \frac{Precision * Recall}{Precision + Recall} \quad (32)$$

$$Accuracy = \frac{TP + TN}{TP + TN + FP + FN} \quad (33)$$

where true positive (TP) is the number of positive instances that are classified as positive, true negative (TN) is the number of negative instances that are classified as negative, false positive (FP) is the number of negative instances that are classified as positive, and false negative (FN) is the number of positive instances that are classified as negative.

## 7 Result and discussion

In this section, the impact of each of the features that introduced in the section3 as Tm-vector component is explained, to show the impact of these features on market predicting accuracy and to determine issues that can be im- proved. 80% of the data is used for model training and 20% for model testing. In data selection training, and testing, the data order is maintained, and a random selection of samples have been prevented. First, Apple stock was selected to extract the best model to report on. Table 4 presents the set of features used in the models and reviewed in Section 6. A description of each of the models and their inputs is given below, and a summary of symbols used in the models is pro- vided in Table 4.

### 7.1 Market sentiment result

In market sentiment analysis, the purpose methods are to use any data set to predict an increase or decrease. In fact, in this example, the analysis is given to each record in the data set using the tag classification of increase or decrease. This analysis is considered an analysis of each tweet or market sentiment. Market sentiment analysis is a finer component and the basis of daily analysis. Several different models were proposed to examine this analysis, which we will discuss in the following:

**Simple IndRNN**: In this model we have only one sample per day. These samples include only market information.

**TM-vector models**: These models include different combinations of our proposed vector. In the following, its var- ious combinations are discussed. In fact, these models are the same TM-vector models based on IndRNN that receive different inputs and try to predict the market direction depending on the input.

1. $TM - vector_M$ : This model creates a vector for each user on behalf of each tweet, which only includes market indicators. The purpose of this model is to examine the impact of market indicators on the stocks.
2. $TM - vector_{FF}$ : The model uses all the features of $So$, $Se$, $Sc$, $Tw$, and $M$. By using this model, we investigate the effect of all the features together on the model prediction result. This model is at the highest level in terms of complexity of input dimensions and includes all textual and numerical data.
3. $TM - vector_{Tw,So,Se,Sc}$ : In this model the features of $Tw, So, Se, Sc$ are used as input features. This model is similar to the $TM - vector_{FF}$ model, except that the $M$ data is omitted from the input data. It has fewer input features and fewer dimensions than the $TM - vector_{FF}$ model.

Table 3: Dataset samples count.

| Stocks | Counts | Positive rate | Negative rate | Unique user |
|---|---|---|---|---|
| Apple | 374856 | 327146 | 47710 | 55082 |
| APX | 9569 | 7895 | 1674 | 163 |
| BA | 66637 | 56153 | 10484 | 641 |
| CAT | 41543 | 34140 | 7403 | 3380 |
| CSCO | 7093 | 6252 | 841 | 187 |
| CVX | 15214 | 13475 | 1739 | 40 |
| DIS | 153032 | 134246 | 18651 | 877 |
| DOW | 14697 | 535 | 125 | 285 |
| GS | 59401 | 50359 | 9042 | 182 |
| HD | 62497 | 55282 | 7139 | 123 |
| IBM | 62871 | 55477 | 7394 | 10440 |
| INTC | 36816 | 31650 | 5166 | 551 |
| JNJ | 42045 | 36070 | 5975 | 448 |
| JPM | 10614 | 9122 | 1492 | 99 |
| KO | 16444 | 14171 | 2273 | 17 |
| MCD | 68810 | 60500 | 8199 | 218 |
| MMM | 11301 | 8946 | 2355 | 47 |
| MRK | 19687 | 17485 | 2202 | 132 |
| MSFT | 156643 | 143562 | 13081 | 979 |
| NKE | 50659 | 44001 | 6653 | 946 |
| PFE | 74733 | 67132 | 7601 | 151 |
| PG | 53866 | 48231 | 5635 | 193 |
| TR | 445 | 3946 | 505 | 318 |

Table 4: Symbols description in the models.

| Parameter Name | Dimension | Features |
|---|---|---|
| So (Social) | $R^6$ | Follower count, friends count, replies, retweets, and favorites |
| Se(sentiment) | $R^3$ | sentiment, subjectivity, and polarity |
| Sc (Score) or user background | $R^4$ | pos user score, neg user score, Author recommendation score, and Representativeness of review author |
| Tw (Tweet) | $R^{Max\ Length*300}$ | text |
| M(Market) | $R^5$ | RSI, MACD, CCI, BB, and MA |
| FF (Full Feature) | $R^n$ | So, Se, SC, Tw, and M |

Table 5: Results of applying the proposed models

| Model (AAPL) | Accuracy | F1 | Recall | Precision |
|---|---|---|---|---|
| IndRNN | 0.69 | 0.80 | 0.93 | 0.71 |
| $TM-vector_{M}$ | 0.88 | 0.92 | 0.87 | 0.97 |
| $TM-vector_{FF}$ | 0.86 | 0.90 | 0.92 | 0.90 |
| $TM-vector_{Tw,So,Se,Sc}$ | 0.87 | 0.91 | 0.85 | 0.99 |
| $TM-vector_{Sc,So,Se}$ | 0.91 | 0.93 | 0.88 | 0.99 |
| $TM-vector_{Tw,So,Se}$ | 0.91 | 0.94 | 0.89 | 0.99 |
| $TM-vector_{So}$ | 0.83 | 0.88 | 0.85 | 0.92 |
| $User2Vec_{Tw,So,Sc}$ | 0.88 | 0.92 | 0.85 | 1.0 |
| $TM-vector_{Tw,Se,So}$ | 0.87 | 0.89 | 0.95 | 0.85 |
| $TM-vector_{Tw,Se,Sc}$ | 0.90 | 0.93 | 0.88 | 0.99 |
| $TM-vector_{Tw,Se}$ | 0.91 | 0.94 | 0.89 | 0.98 |
| $TM-vector_{So,Se,Sc}$ | 0.68 | 0.81 | 0.68 | 0.1 |
| $TM-vector_{Tw}$ | 0.69 | 0.80 | 0.93 | 0.71 |
| $TM-vector_{M,So,Se}$ | 0.9500 | 0.92 | 0.91 | 0.93 |
| V | 1 | | | |
| UNH | 15576 | 13296 | 2280 | 80 |
| UTX | 9977 | 8654 | 1323 | 267 |
| V | 34638 | 29494 | 5144 | 163 |
| VZ | 18913 | 16400 | 2513 | 116 |
| WBA | 29835 | 26496 | 339 | 256 |
| WMT | 13268 | 11562 | 1706 | 313 |
| XOM | 23439 | 19881 | 3558 | 250 |

4. $TM{-}vector_{Sc,So,Se}$: In this model the features of $Sc$, $So$, and $Se$ are considered as inputs and the text and $M$ features are ignored. The purpose of this model is to present a model based on numerical data and to investigate the effect of numerical data on prediction.
5. $TM - vector_{Tw,So,Se}$:This model attempts to predict using only features extracted from the Twitter and the features of $M$ and $Sc$. Stock index information that could be useful information is not included in this model.
6. $TM{-}vector_{So}$:In this model, only $So$ data, is used to predict stocks. The purpose of the model is to predict through $So$ data, which contains useful information about users who have commented on the stock.
7. $TM - vector_{Tw,So,Sc}$ : The input to this model is $Tw$, $So$, and $Sc$ data, which measures the impact of textual data and extracted manual features. The $TM - vector_{Tw,So,Se}$ and $TM - vector_{Tw,Se,Sc}$ are similar to this model except that the $Se$ is replaced by $Sc$ and $So$.
8. $TM{-}vector_{Se,So,Sc}$ : This model is another of models that considers only numerical features for prediction. At the input of this model, $M$ data is omitted.
9. $TM - vector_{Tw}$:This model only considers textual data generated on Twitter as input and excludes other features.
10. $TM - vector_{M,So,Se}$:In this model the features of $M$, $So$, and $Se$ are considered as inputs and the text are ignored.

According to table 5 the best results include M, So, Sc features, which is 95% of accuracy for the Apple stock. From the results we can conclude that the most accurate improvement is related to the selection of the TM-vector model, which increases the number of samples from the number of days to the number of tweets. In addition, the effect of the individual features was investigated, which show that the tweets did not increase the accuracy of the model, and that the user rating alone could not improve the accuracy, but improves accuracy when uses with $So$ or $Se$. Another out coming is that market data has the most impact on accuracy than other features. The best model based on the results which is $TM - vector_{M,So,Se}$, so that we could use it for the other stocks.

### 7.2 TM-vector for DOW 30 stocks

Unfortunately, in this article, due to the lack of reference to the data collection period and also the lack of publication of the used data, it is not possible to compare the experimental results with other approaches in the literatures. In order to investigate the effect of the proposed vectors and properties, in addition to the IndRNN model, we also tested two models, LSTM and GRU. These three models were also considered to be exactly equal in terms of structure and number of implementation steps. Each of the IndRNN, LSTM, and GRU models has hyper-parameter that are mentioned in Table 6. These values are important for reconstructing the results. The effect of batch size is also shown in Figures 6 and 7 per each stock. The results of all stocks are presented in table 7. All results are reported by the selected $TM-vector_{M,So,Se}$ model. The most accurate of the Dow 30 stocks listed is WMT. Other stocks have obtained the accuracy More than 75%. Some stocks have less data because of the volume of daily trades of them and as a result less comments from users on Twitter. This shortcoming has led to unsatisfactory results for some stacks such as MMM and V. The analysis of the misclassified data, it can be concluding that the lack of collected data and the lack of a suitable score for each user results in to this misclassification. Another analysis is performed on poor classified stocks, which blames on insufficient data for these stocks. Most approaches in the literature only cover a subset of the stocks examined in this pa- per. In this regard, no comparisons can be made between approaches due to the unavailability of their data, the different data collection intervals, and the lack of focus on all stocks examined in this study.

### 7.3 Error analysis

A detailed error analysis of the performance of the proposed TM-vector (IndRNN) model is performed in this section. Figure 5 reports the set of confusion matrices for the 30 stocks examined in this work. In general, it can be seen that the proposed strategy tends to predict decreases in stock prices (negative samples). In-depth analysis of some of the incorrect predictions led us to conclude that due to the proximity of the incremental sample vectors (positive samples) to the subtractive sample vectors (negative samples) they resulted in incorrect classification. Also, the imbalance of be- tween positive and negative classes has caused this problem. The issue is also true for WBA, Gs, IBM, KO, and MRK stocks because the lack of negative samples makes the model unable to learn a good representation of the difference be- tween the two classes and to perform at least weaker than the class samples. One of the future things that can be done to reduce errors is to collect more data.

Selected models generally omit contextual data, but there are a number of approaches to improve text-based models that can be useful. In manual and statistical study of 100 re- views that were mistakenly classified, we conclude that for most part, the negation scope has led to the incorrect classification of these reviews. As part of our future work, our goal is to apply negation scope as a manual feature along with the features selected automatically by the proposed approach. In (65), the author used 19 negative words with 90 patterns to detect the negation scope which led to increase in the accuracy of their classifier. Similar patterns can be used to improve the proposed approaches.

In (66) a kind of neural network called the" CapsuleNet". In this network, along with the activation of low-level neuron that shows the likelihood of detecting specific feature, sends the vector of the attributes to the next layer. Depending on the type of data (here text), these features can be the length of the words, the position of words, the length of the negation scope, and etc. Another benefit of this network is that, un- like CNNs, less data is required for training. CapsuleNet is another approach that can be applied to current work. We expect this network to get better results than investigated approaches.

Table 6: The models hyper-parameters.

| - | IndRNN | LSTM | GRU |
|---|---|---|---|
| Of epoch | 100 | 100 | 100 |
| Of layer | 2 | 2 | 2 |
| Of block | 14 | 14 | 14 |
| Learning rate | 0.001 | 0.001 | 0.001 |
| Activation function | sigmoid | sigmoid | sigmoid |
| Recurrent dropout | 0.5 | 0.5 | 0.5 |
| L2 regularization | 0.0001 | 0.0001 | 0.0001 |
| Dropout | 0.5 | 0.5 | 0.5 |

### 7.4 Daily Analysis

In the case of classification at the tweet level, model analysis can not provide a view of better models. For this purpose, we use daily based analysis for each model. To get the accuracy of the models in daily mode, we use the following rule:

$$Daily_{Predict} = \begin{cases} Pos, if\ count(Pos) > count(Neg) \\ Neg, if\ count(Pos) > count(Neg) \\ Neg, Otherwise \end{cases} \quad (34)$$

Count (Pos)and count (Neg) determines the number of pre- dictions per day in positive and negative cases respectively. If these two values are equal to each other, the strategy of the proposed model is to consider the state negative. In fact, in this case, this solution is presented to avoid a 50-50 decision strategy. Table 8 shows the results for Apple in all data combinations for the three models TM-vector (IndRNN), TM- vector (LSTM), and TM-vector (GRU). In the data mode Tm- vector(M) the TM-vector (LSTM) approach achieved the highest accuracy of 0.5856, while the proposed IndRNN method achieved the lowest accuracy of 0.5612. In the TM-Vector (Txt) model, the TM-vector (IndRNN) approach was able to reach an accuracy of 0.5710 compared to the other two methods on this data. However, compared to the TM-vector (LSTM) on the data market M, TM-vector (IndRNN) is still at lower accuracy.
The combination of social and sentiment data was somewhat more effective in TM-Vector (So; Se). TM-vector (IndRNN) on data combination (So; Se) achieved an accuracy of 0.6809. TM-vector (IndRNN) on combining market data with so in TM-Vector mode (M; So) and combining with data SE in TM-Vector (M; Se) achieved an accuracy of 0.69 and achieved the best result among algorithms compared to equal inputs. The TM-vector (IndRNN) on TM-Vector (M; So; Se) was able to achieve the highest accuracy of 0.7610 among all algorithms and data. Daily results showed that the distribution of features could affect the accuracy of the models, and practically the raw text of the tweets has no effect on increasing the accuracy of the models and can be filtered. Also, market data alone can not get a daily forecast view in any of the proposed models.

### 7.5 Sensitivity Analysis

For sensitivity analysis we study the effectiveness of the batch size in the TM-vector (IndRNN), LSTM, and, GRU model. The batch size represents the number of data samples that is propagated through the network. The larger batch consumes more memory, and the smaller batch increase the training time. For this purpose, we tested different values of 128, 256, 512, 1024, 2048, and 4096 for batch size. Figure 6 and 7 show the obtained result by different batch size.

Table7: The results of experiments performed on Dow 30 by GRU, LSTM, and IndRNN models.

| | **TM-vector (IndRNN)** | | | | **TM-vector (LSTM)** | | | | **TM-vector (GRU)** | | |
|---|---|---|---|---|---|---|---|---|---|---|---|
| | Acc | Prec | Rec | F1 | Acc | Prec | Rec | F1 | Acc | Prec | Rec |
| Apple | **0.9500** | 0.9248 | 0.9103 | 0.9398 | 0.8047 | 0.8563 | 0.8545 | 0.8551 | 0.9083 | 0.9376 | 0.9261 |
| AXP | **0.8393** | 0.8906 | 0.8858 | 0.8881 | 0.7158 | 0.7699 | 0.8538 | 0.8092 | 0.7461 | 0.8236 | 0.8158 |
| BA | **0.8980** | 0.7464 | 0.8360 | 0.7886 | 0.7656 | 0.8078 | 0.7619 | 0.7836 | 0.8920 | 0.8924 | 0.9175 |
| CAT | **0.9347** | 0.9422 | 0.9683 | 0.9550 | 0.9201 | 0.9433 | 0.9614 | 0.9522 | 0.8552 | 0.8557 | 0.8911 |
| CSCO | **0.8586** | 0.8549 | 0.9334 | 0.8924 | 0.7984 | 0.8279 | 0.8078 | 0.8175 | 0.7618 | 0.7887 | 0.7881 |
| CVX | **0.8474** | 0.8461 | 0.9555 | 0.8974 | 0.5447 | 0.5571 | 0.7130 | 0.6379 | 0.7289 | 0.7945 | 0.7988 |
| DIC | **0.8564** | 0.8767 | 0.8318 | 0.8536 | 0.7257 | 0.7869 | 0.7828 | 0.7843 | 0.8466 | 0.8739 | 0.8881 |
| DOW | **0.9470** | 0.9293 | 1.0 | 0.9633 | 0.7257 | 0.7869 | 0.7828 | 0.7843 | 0.9242 | 0.9238 | 0.9798 |
| GS | **0.7917** | 0.8882 | 0.7497 | 0.8130 | 0.8230 | 0.8813 | 0.8410 | 0.8603 | 0.7665 | 0.8319 | 0.8032 |
| HD | **0.8542** | 0.8817 | 0.8946 | 0.8881 | 0.8230 | 0.8813 | 0.8410 | 0.8603 | 0.7994 | 0.8270 | 0.8728 |
| IBM | **0.8124** | 0.9116 | 0.7633 | 0.8308 | 0.7519 | 0.7941 | 0.8231 | 0.8078 | 0.7602 | 0.8215 | 0.7962 |
| INTC | **0.8406** | 0.8233 | 0.9315 | 0.8740 | 0.7996 | 0.8822 | 0.8104 | 0.8445 | 0.7998 | 0.8499 | 0.8528 |
| JNJ | **0.8626** | 0.9012 | 0.8923 | 0.8967 | 0.7989 | 0.8699 | 0.8319 | 0.8501 | 0.7951 | 0.8410 | 0.8655 |
| JPM | **0.8605** | 0.8689 | 0.9535 | 0.9092 | 0.8497 | 0.8823 | 0.8948 | 0.8886 | 0.8347 | 0.8871 | 0.8642 |
| KO | **0.8465** | 0.9203 | 0.8398 | 0.8782 | 0.6817 | 0.7175 | 0.8365 | 0.7721 | 0.7507 | 0.8109 | 0.8010 |
| MCD | **0.8164** | 0.8816 | 0.8220 | 0.8507 | 0.6623 | 0.7735 | 0.6966 | 0.7326 | 0.7537 | 0.7925 | 0.8533 |
| MMM | **0.8055** | 0.7207 | 0.8415 | 0.7764 | 0.6581 | 0.6781 | 0.7535 | 0.7132 | 0.6953 | 0.7226 | 0.7494 |
| MRK | **0.8362** | 0.9022 | 0.8015 | 0.8488 | 0.7374 | 0.7900 | 0.8312 | 0.8097 | 0.8078 | 0.8256 | 0.9070 |
| MSFT | **0.8437** | 0.8940 | 0.8382 | 0.8652 | 0.6190 | 0.8050 | 0.5953 | 0.6837 | 0.8395 | 0.8597 | 0.9189 |
| NKE | **0.8348** | 0.8859 | 0.8669 | 0.8762 | 0.7712 | 0.8397 | 0.8254 | 0.8321 | 0.7664 | 0.8067 | 0,8691 |
| PFE | **0.8129** | 0.8747 | 0.8179 | 0.8453 | 0.7532 | 0.8334 | 0.7722 | 0.8013 | 0.7300 | 0.7913 | 0.7908 |
| PG | **0.8056** | 0.8739 | 0.8109 | 0.8412 | 0.6802 | 0.7623 | 0.7364 | 0.7485 | 0.7293 | 0.7646 | 0.8418 |
| TRV | **0.8886** | 0.8785 | 0.9284 | 0.9027 | 0.7329 | 0.7668 | 0.8453 | 0.8037 | 0.7138 | 0.7794 | 0.8147 |
| UNH | **0.8739** | 0.8860 | 0.9045 | 0.8951 | 0.8065 | 0.8197 | 0.8737 | 0.8454 | 0.8460 | 0.8593 | 0.8938 |
| UTX | **0.8441** | 0.8592 | 0.9354 | 0.8956 | 0.8161 | 0.8441 | 0.8551 | 0.8492 | 0.7916 | 0.8585 | 0.8048 |
| V | **0.8085** | 0.8988 | 0.8136 | 0.8540 | 0.7094 | 0.7635 | 0.8314 | 0.7955 | 0.7494 | 0.8293 | 0.7953 |
| VZ | **0.8943** | 0.9292 | 0.9047 | 0.9167 | 0.8091 | 0.8351 | 0.8730 | 0.8532 | 0.8583 | 0.8823 | 0.8974 |
| WBA | **0.7952** | 0.8850 | 0.7578 | 0.8164 | 0.7436 | 0.7850 | 0.7890 | 0.7865 | 0.7379 | 0.7863 | 0.7743 |
| WMT | **0.9470** | 0.9525 | 0.9683 | 0.9603 | 0.8451 | 0.9007 | 0.8805 | 0.8903 | 0.8161 | 0.8441 | 0.8551 |
| XOM | **0.8316** | 0.8336 | 0.9190 | 0.8742 | 0.7039 | 0.7693 | 0.7928 | 0.7805 | 0.8161 | 0.8441 | 0.8551 |

Table8: The results of experiments performed on Apple by GRU, LSTM, and IndRNN models per day.

| | **TM-vector (IndRNN)** | | | | **TM-vector (LSTM)** | | | | **TM-vector (GRU)** | | | |
|---|---|---|---|---|---|---|---|---|---|---|---|---|
| **Model (AAP)** | **Acc** | **Prec** | **Rec** | **F1** | **Acc** | **Prec** | **Recall** | **F1** | **Acc** | **Prec** | **Rec** | **F1** |
| **TM-Vector (M)** | 0.5612 | 0.6862 | 0.6899 | 0.6880 | 0.5856 | 0.6830 | 0.7003 | 0.6915 | 0.5710 | 0.5829 | 0.5890 | 0.5859 |
| **TM-Vector (Txt)** | 0.5710 | 0.5942 | 0.6648 | 0.6275 | 0.5391 | 0.5441 | 0.5561 | 0.5500 | 0.5233 | 0.5458 | 0.5489 | 0.5473 |
| **TM-Vector (So; Se)** | 0.6809 | 0.6732 | 0.6801 | 0.6766 | 0.6728 | 0.6701 | 0.6712 | 0.6706 | 0.6659 | 0.6831 | 0.6890 | 0.6860 |
| **TM-Vector (M; So; Se)** | 0.7610 | 0.7743 | 0.7842 | 0.7792 | 0.6221 | 0.6688 | 0.6790 | 0.6738 | 0.6621 | 0.6730 | 0.6807 | 0.6768 |
| **TM-Vector (M; So)** | 0.6919 | 0.7025 | 0.7129 | 0.7076 | 0.6911 | 0.6914 | 0.7308 | 0.7105 | 0.6810 | 0.6721 | 0.6795 | 0.6757 |
| **TM-Vector (M; Se)** | 0.6920 | 0.7017 | 0.7234 | 0.7123 | 0.5790 | 0.6130 | 0.6557 | 0.6336 | 0.6104 | 0.6184 | 0.6301 | 0.6241 |

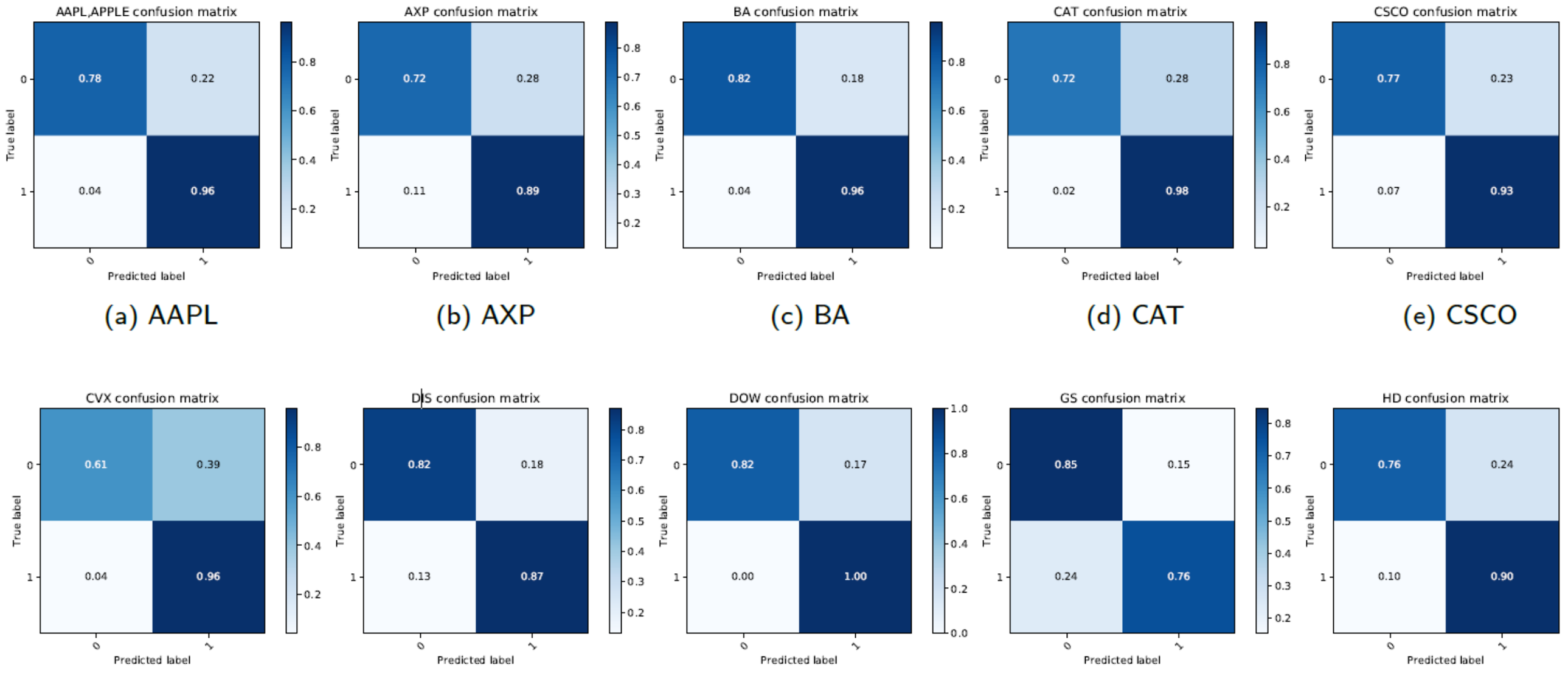

(a) AAPL (b) AXP (c) BA (d) CAT (e) CSCO

(f) CVX (g) DIS (h) DOW (i) GS (j) HD

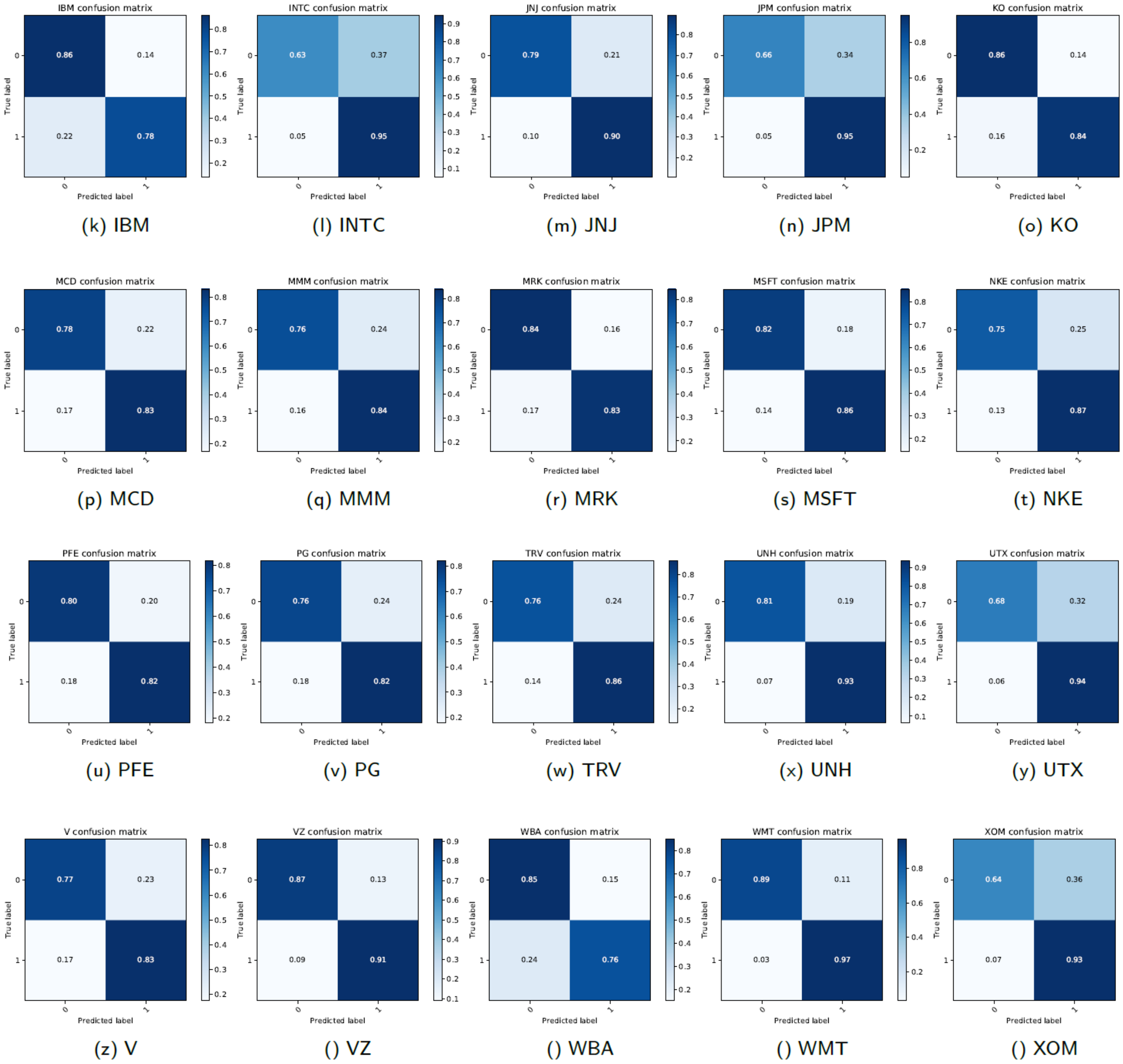

Figure 5: Confusion matrices for the predictions on the Dow 30 using the proposed IndRNN model.

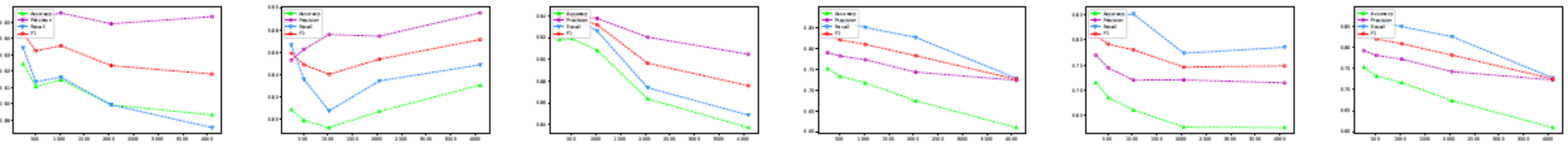

(a) Apple IndRNN (b) Apple LSTM (c) Apple GRU (d) AXP IndRNN (e) AXP LSTM (f) AXP GRU

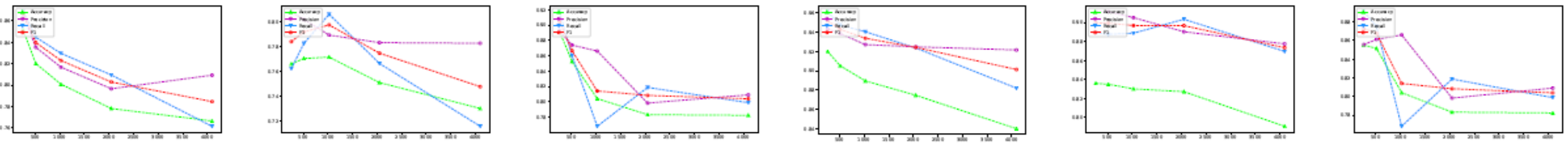

(g) BA IndRNN (h) BA LSTM (i) BA GRU (j) CAT IndRNN (k) CAT LSTM (l) CAT GRU

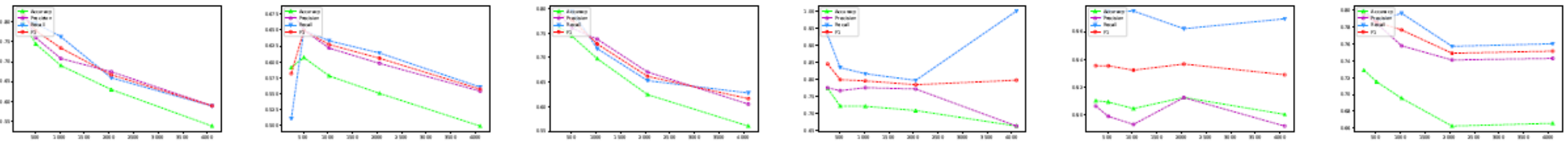

(m) CSCO IndRNN (n) CSCO LSTM (o) CSCO GRU (p) CVX IndRNN (q) CVX LSTM (r) CVX GRU

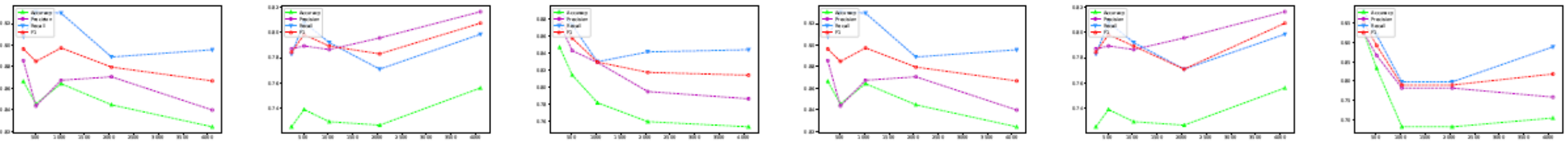

(s) DIS IndRNN (t) DIS LSTM (u) DIS GRU (v) Dow IndRNN (w) Dow LSTM (x) Dow GRU

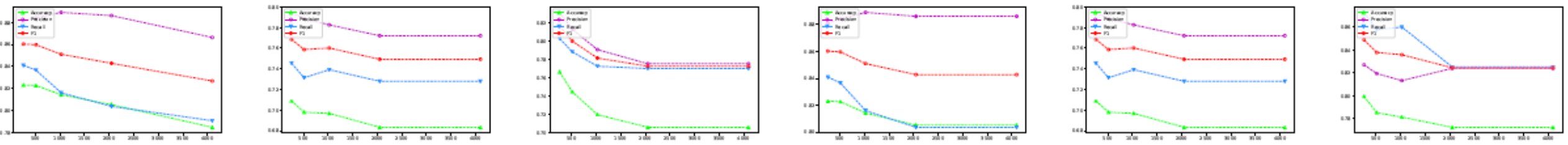

(y) GS IndRNN (z) GS LSTM () GS GRU () HD IndRNN () HD LSTM () HD GRU

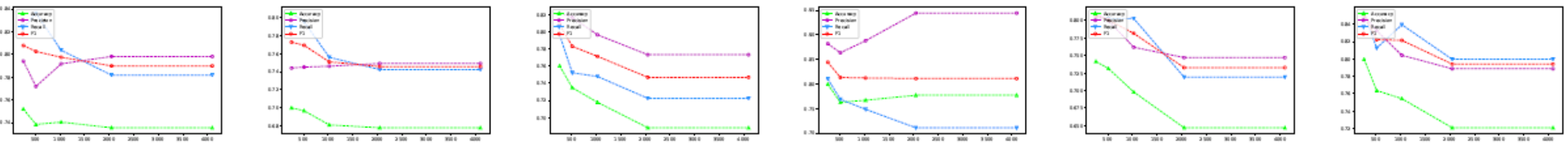

() IBM IndRNN () IBM LSTM () IBM GRU () INTC IndRNN () INTC LSTM () INTC GRU

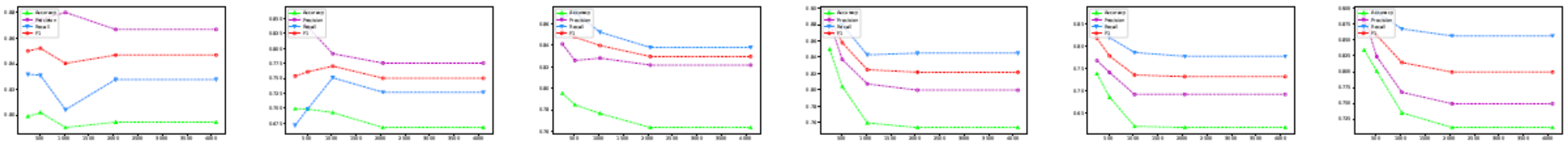

() JNJ IndRNN () JNJ LSTM () JNJ GRU () JPM IndRNN () JPM LSTM () JPM GRU

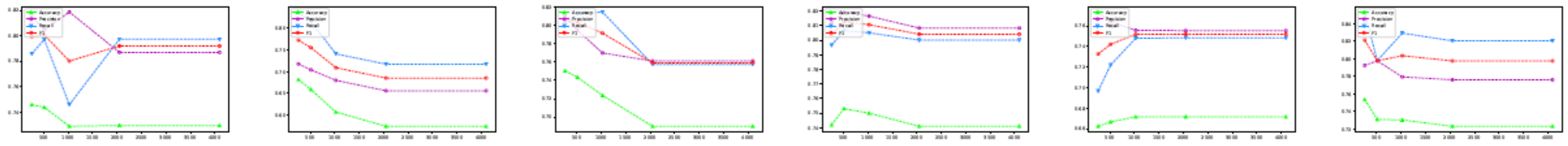

() KO IndRNN　() KO LSTM　() KO GRU　() MCD IndRNN　() MCD LSTM　() MCD GRU

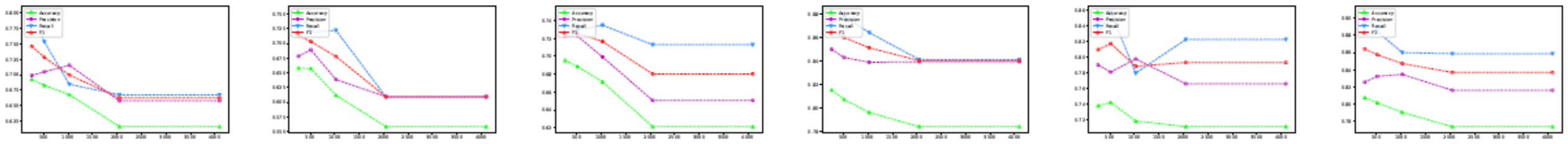

() MMM IndRNN　() MMM LSTM　() MMM GRU　() MRK IndRNN　() MRK LSTM　() MRK GRU

Figure 6: The results of experiments performed on Dow 30 to fine tuning the batch-size GRU, LSTM, and IndRNN models.

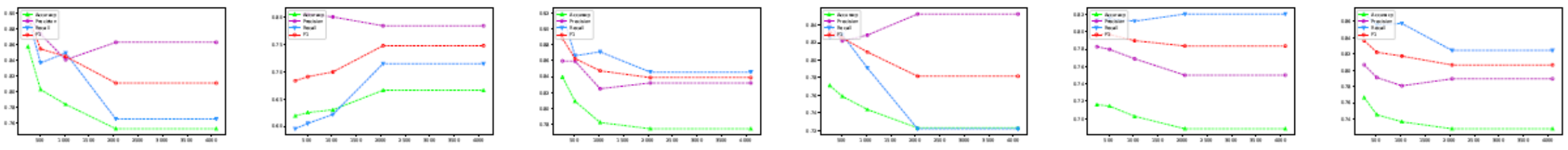

(a) MSFT IndRNN　(b) MSFT LSTM　(c) MSFT GRU　(d) NKE IndRNN　(e) NKE LSTM　(f) NKE GRU

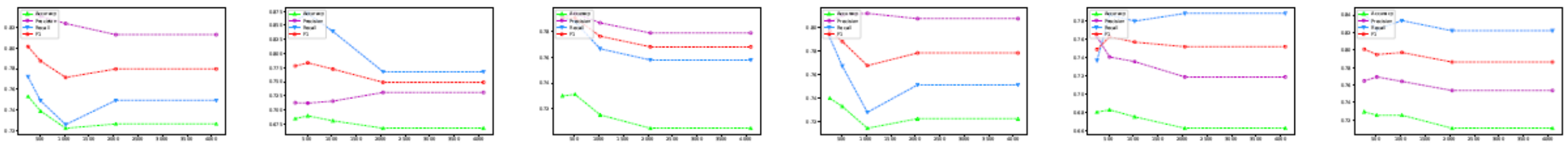

(g) PFE IndRNN　(h) PFE LSTM　(i) PFE GRU　(j) PG IndRNN　(k) PG LSTM　(l) PG GRU

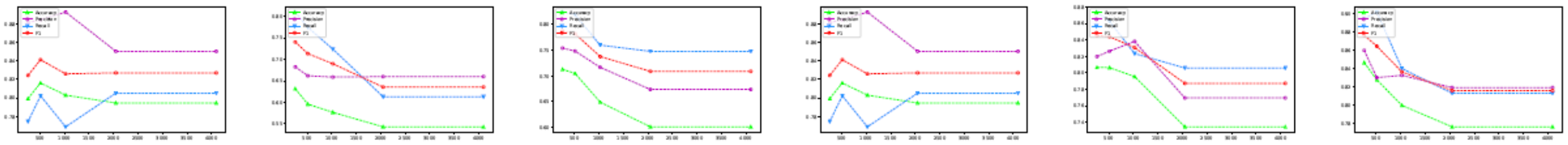

(m) TRV IndRNN　(n) TRV LSTM　(o) TRV GRU　(p) UNH IndRNN　(q) UNH LSTM　(r) UNH GRU

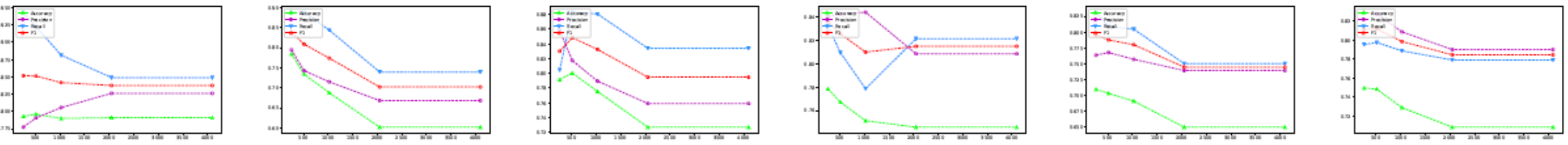

(s) UTX IndRNN　(t) UTX LSTM　(u) UTX GRU　(v) V IndRNN　(w) V LSTM　(x) V GRU

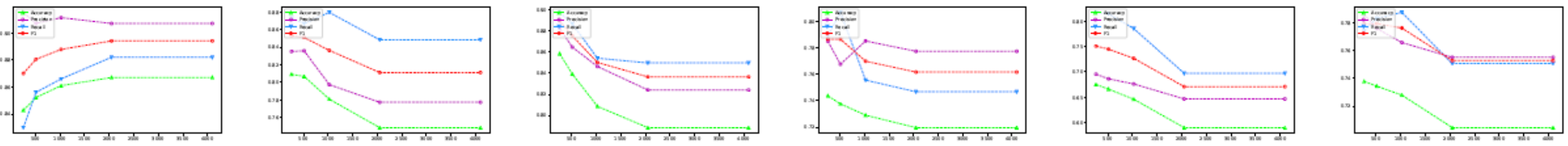

(y) VZ IndRNN　(z) VZ LSTM　() VZ GRU　() WBA IndRNN　() WBA LSTM　() WBA GRU

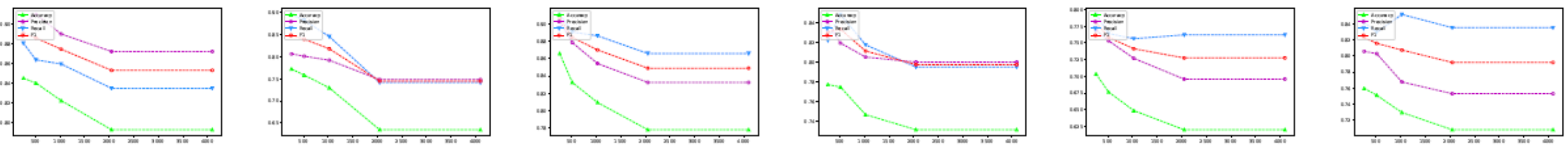

() WMT IndRNN　() WMT LSTM　() WMT GRU　() XOM IndRNN　() XOM LSTM　() XOM GRU

Figure 7: The results of experiments performed on Dow 30 to fine tuning the batch-size GRU, LSTM, and IndRNN models.

## 8 Conclusion and Future work

Predicting the stock market's direction is one of the most important issues in which deep learning and machine learning can play a crucial role. The most important advantage of deep learning over machine learning is the automatic selection of features. In market direction forecasting, two classes called increase or decrease are usually considered the target of the forecast, and models are trained for these two classes. Various approaches for this purpose have been proposed in the literature. Approaches trained only on market or social data increase the likelihood of overfitting and underfitting. On the other hand, users' social information can be considered essential for forecasting. In the proposed approach from three different data entitled 1) market data that included statistical indicators, 2) Twitter data that included raw text and other social features of users and 3) emotion analysis features that for each tweet specifying its polarity was used. The proposed TM-vector (IndRNN) approach is a recurrent- based network approach that attempts to model data as independent dependencies. IndRNN blocks consider the temporal influence between input features. Depending on the type of data, this approach achieved a maximum accuracy of 95% in the analysis of market sentiment and an accuracy of 76% in the daily analysis of Apple stock. The main contribution of the proposed approach compared to other models presented in the literature of related research is the simulation of Twitter and market data which have not been considered simultaneously. According to the results of the evaluation, it was shown that this data alone does not contain property information and can not provide a better view of the future of a stock. The proposed approach, like all forecasting models, is not without limitations. One of the weak points of this approach is that its decision is very sensitive to numbers. In TM-vector (IndRNN), if $count(Pos) = N + 1$ and $count(Neg) = N$ then the class is considered positive. Also, if the next day's price is only a few cents more than the price of today, it is considered positive. Future studies can focus on the heuristic threshold for solving these two issues.